\documentclass{kluwer}    % Specifies the document style.
\usepackage[]{graphicx}
\newdisplay{guess}{Conjecture}

\begin{document}                                                                                   
\begin{article}
\begin{opening}         
\title{Distances on Cosmological Scales with VLTI} 
\author{Margarita \surname{Karovska}, Martin \surname{Elvis}, and Massimo \surname{Marengo}} 
\runningauthor{Margarita Karovska}
\runningtitle{Distances on Cosmological Scales with VLTI}
\institute{Harvard-Smithsonian Center for Astrophysics}
%\date{April 15, 1993}

\begin{abstract}

We present here a new method using interferometric measurements 
of quasars, that allows the determination of direct geometrical
distances on cosmic scales. 
Quasar Broad Emission Line Regions sizes
provide a `meter rule' with which to measure the metric of the
Universe. 
This method is less dependent of 
model assumptions, and even of variations in the fundamental 
constants (other than c).
We discuss the spectral and spatial requirements on the VLTI
observations needed to carry out these measurements.

\end{abstract}
%\keywords{sample, \LaTeX}

\end{opening}           

\section{Introduction}  

%Accurate measurements of distances to astronomical objects provide 
%crucial information on the
%luminosities
%and ages of various sources as well as the distance scale.
Determining the 3-D geometry of the Universe involves
 measurement of the large ``cosmic''  scale
distances of high redshift sources such as distant supernovae and quasars. 
Cosmic distance scale determination methods include
relative and absolute distance estimators.
Absolute distance estimators include various applications of the:
 Baade-Wesselink method (eg. Sasselov and Karovska 1994; Marengo et al., this workshop), 
and distances estimators using time delays (eg. SN1987A; Panagia et
 al. 2001).
Relative distance estimating methods include 
period-luminosity relation for Cepheid star variability (used by the {\em Hubble} 
Key Project ; Freedman et al. 2001), and supernovae
of type SN1a as 'standard candles'. Most strikingly,
recent distance measurements using the brightness of supernovae
of type SN1a at z$\sim$1.5 as 'standard candles' suggest that that the universe has a non-zero cosmological constant,
$\Lambda$, so that the expansion of the Universe is currently
accelerating (Perlmutter et al. 1999).

Relative distance estimators often involve assumptions and correlations,
and have inevitable model
dependencies.
Absolute methods on the other hand 
have the advantage of having lesser dependence on physical models
and provide an independent way to determine the distance scale.

We describe here an absolute method for estimating distances to quasars at different
{\it z} using a time delay measurements of linear sizes of quasars Broad
Emission Line Regions (BELRs, Peterson 1997) and
measurements of their angular sizes 
 using large baseline interferometers such as the VLTI.
This method could allow determining 
direct geometrical distances to
radio-quiet quasars and therefore 
measuring the cosmological constant,
$\Lambda$, with minimal assumptions.

\section{Quasar Parallax}

\begin{figure}\label{f1}
\centerline{\includegraphics[width=26pc,angle=0]{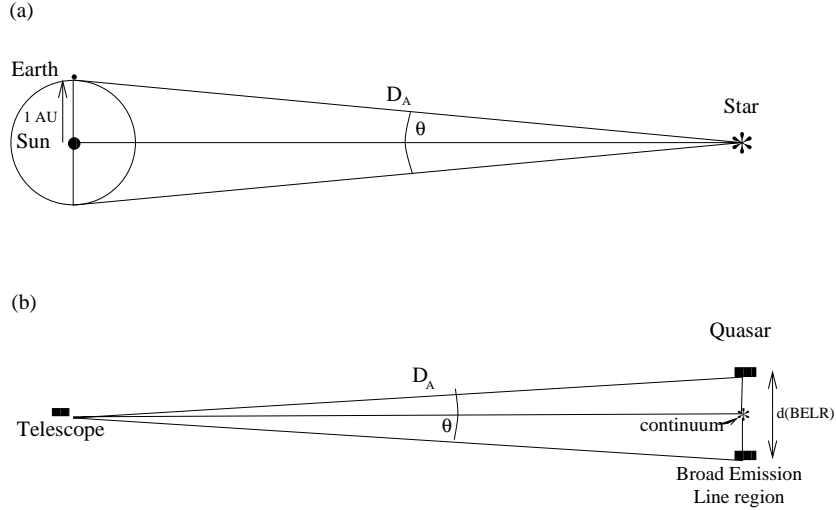}}
\caption{Measuring geometric distances to astronomical objects:
(a) parallax to a star; (b) broad emission line region distance
to a quasar.}
\end{figure}

We propose to use a new geometric method to determine
distances to quasars that is equivalent to the parallax
method. However, in this case instead of
using the Earth's orbit as a known `standard length' (Figure 1a), we
use the size of the quasar BELR (Figure 1b) as measured using reverberation
mapping and interferometry (Elvis and Karovska
2002).

The
size of the BELR is known from light travel time
measurements.
BELRs emit a series of strong
lines from permitted transitions,
indicating Doppler velocities of
5000~--~10,000~km~s$^{-1}$ --- defining characteristic of quasars. 
The Broad Emission Lines (BELs) respond to changes in the continuum
source in the center of the quasar by changing their intensity with a time-lag of a few days
or more (for higher luminosity and redshift). 
The time-lag is induced by the light travel time from the
continuum source in the center of the quasar to the gas clouds in the
BELRs that produce the broad
emission lines (Peterson 1993, 2001).
By studying the response
of the BELs to continuum changes it is possible to map out the
structure of the region emitting the BELs. This method is called
`reverberation mapping'.

Reverberation mapping exists for a few
tens of active galaxies and quasars. The size of the broad
emission line region depends on the line measured.
Observations indicate that a single timescale  
dominate for each emission line (Nezter \& Peterson 1997).
For low redshift quasars the size of the BELRs is $\sim$10~ light days.
The timescale of BEL variations increases with redshift as
$(1+z)^{2.5}$, since not only does the BELR size increase, but
also the intrinsic variation timescale is dilated by $(1+z)$.
Typical reverberation time for quasars
at z=3 requires observing programs spanning of 3-5~years.

The `angular diameter distance' derived using the reverberation mapping and the
interferometric measurements can be mapped as a function of
redshift. This relation will describe the basic characteristic of
the space-time metric of the Universe, and could eventually allow 
a derivation of $\Lambda$
(Elvis and Karovska 2002).  
This method of derivating $\Lambda$ is much less dependent on physical 
models and of
changes in the fundamental constants (other than $c$, the speed
of light) because it uses a `standard length' approach rather than a
`standard candle'.

\section{Interferometry of Quasar BELRs}

Measurements of the angular sizes of BELRs are not easy
even for the nearby quasars
and AGNs.
For example, at z$<$0.2 the BELRs have measured lags that imply angular sizes
$\sim$0.1~mas (Fig 2.). This angular size is about 10 times smaller then
the baselines of the existing ground based interferometers operating in
optical and near-IR wavelengths (e.g. NPOI, IOTA). In addition, their
sensitivity and the limiting magnitude are far bellow what is
required for measuring the BELRs.

\begin{figure}\label{f2}
\centerline{\includegraphics[width=24pc,angle=0]{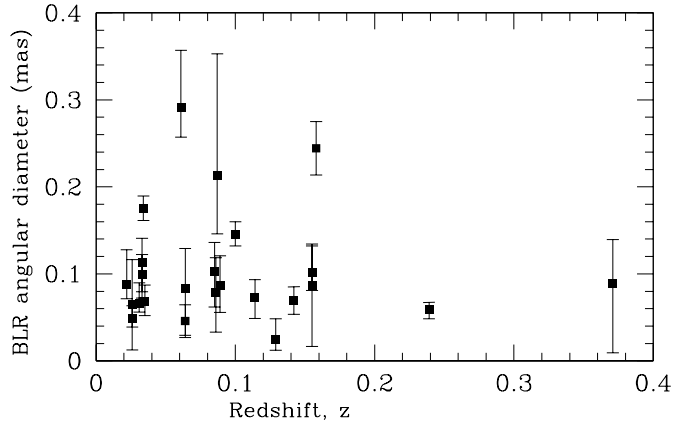}}
\caption{Angular diameters for the H$\alpha$ and H$\beta$ BELRs
of nearby active galaxies, assuming
H$_0$=65~km~s$^{-1}$Mpc$^{-1}$ (Peterson et al. 1998, Kaspi et
al. 2000). (The values of $\Omega$ and $\Lambda$ are unimportant
at these low redshifts.}
\end{figure}

To measure accurately the sub-milliarcsecond scale BELR angular
diameters
therefore
requires very long-baseline interferometers with large aperture
telescopes, and with a capability of observing in selected quasar emission lines.
In addition, the
baseline of an interferometer that can resolve the BELR increases
with $z$ for a fixed physical size, because
the
observed wavelength of any emission line increases as $(1+z)$.

Reverberation mapping suggests that shorter
wavelength emission lines 
(in the UV or X-ray bands) may be more
favorable for BELR size measurements. However,
the space-based interferometers operating at sub-optical
wavelength range are currently only in planning stages, and the
ground-based interferometry is at present the only mean to carry out these
measurements.

Several long-baseline ground-based optical and near-IR
interferometers that have a
potential of measuring the sizes of the BELRs in nearby quasars and
AGNs are currently being built or commissioned.
These interferometers, including the VLTI,
are employing large aperture telescopes (e.g. VLTI 1.8m to 8-m) on
baselines of several hundred meters and have the sensitivity required 
to observe BELRs
in selected spectral lines used in the reverberation measurements.

Since lag-time and angular size
measurements must be carried out using the same emission
lines,
we are currently exploring optical and near-IR lines that may be
suitable for reverberation mapping studies as
well as for interferometry.
Our initial studies indicate that 
in the optical H$\beta$, H$\alpha$ lines could be appropriate
for these measurements.
The next generation VLTI instruments operating in the optical
would be able to resolve the
nearby AGN BELRs and study their geometry.
This is very important, because the geometry of quasar BELRs is not well known. Possibilities range
from a simple orderly wind (Elvis 2000) to a maximally chaotic
distribution of clouds (Baldwin et al. 1995).

In the near-IR the VLTI could measure the BELRs sizes of nearby (very
low redshift) unobscured AGNs (eg. with AMBER; Petrov et al. 2000).
Possible useful lines include HeII 10830\AA, H Pa-$\alpha$ (18751\AA),
and H Pa-$\beta$ (12818\AA).
For the nearest AGNs (eg. Cen A) the expected BELRs sizes are significantly larger and therefore one could 
carry out imaging and study their geometry (Paresce 2001).
For more distant AGNs and quasars one can then use this information to
estimate the size of their BELRs.
Our simulations show that using high-signal to noise visibility
measurements (less then 1\%) and a model
brightness distribution, one can significantly exceed the limiting
resolution
of the VLTI and estimate diameters much smaller then 1-2 mas.
By assuming a model for the geometry of the BELRs
the VLTI observations could therefore determine the distances to the nearby AGNs
and quasars, and improve the $H_0$ accuracy to $\sim$5\%.

However, there are a number of issues regarding the feasibility of the
interferometric measurements that need to be addressed. 
For example, there is
an inherent problem in relating reverberation lag times as
currently measured to the angular sizes that an interferometer
would measure. Even perfectly sampled data will produce a
broadened cross-correlation function (CCF), and it is not obvious
which delay time, e.g. the CCF peak, or the CCF centroid,
corresponds to the angular size from an image.  
The reverberation mapping measures a
`responsivity weighted' size of the light delay surface, while
interferometry measures an `emission weighted' size. 

The amplitude of
variation of the emission lines is much less than that of the
continuum (eg. factor 2 changes in the UV continuum
produce only 20\% changes in the emission lines)
%2001, figure~31). The high ionization and high order lines vary
%most strongly, but they also tend to be weak (Ulrich et al.,
%1991).
By 
monitoring a number of
quasars for continuum changes, one could identify the
onset of a significant continuum increase in a particular quasar,
which will then trigger
`target of opportunity' VLTI measurements.
The VLTI could observe the
target quasar in both low and high
continuum states, with appropriate lags included, so that the
difference map will be a measure only of the `responsive
fraction' of the line, i.e.  that part which responds to
continuum changes. 
Evidently, the best approach would be to continuously monitor
quasars with the VLTI and derive the lag times and sizes
from the same data.

\section{Conclusions}

Quasar BELRs  provide a meter
rule for determining the cosmic distance scale independent
of changes in the fundamental constants
(other than $c$, the speed of light). We describe a new absolute distance
estimating method
using reverberation time-lag measurements of the linear BELR sizes,
combined with the VLTI interferometric angular size measurements, to
determine the distances to quasars. 
VLTI can measure the sizes of BELRs and explore their geometry at low
redshifts. This will
provide input to BELR models and allow measurement of $H_0$.
Long-baseline interferometers with a resolution of 0.01~mas are needed to
measure the size of the BELR in z=2 quasars, and therefore the
3-D geometry of the Universe (and $\Lambda$) which appear
plausible with near future extrapolations of current
technology.

\acknowledgements

This work was supported in part by NASA contract
NAS8-39073 (Chandra X-ray Center).

\end{article}
\end{document}